\documentclass{appolb}%
\usepackage{graphicx}
\usepackage{amssymb}
\usepackage{amsmath}
\usepackage[colorlinks=true,linktocpage=true,linkcolor=blue,citecolor=blue]{hyperref}
\usepackage[usenames,dvipsnames]{color}
\usepackage{bm}  
\usepackage{amsfonts}

\begin{document}
\renewcommand*{\thefootnote}{\fnsymbol{footnote}}

\title{Geometric scaling and $C$-odd exchanges \\ at low and high energies}
\author{Micha{\l} Prasza{\l}owicz
\thanks{Presented at ``Diffraction and Low-$x$ 2026'', Kalamata (Greece), Sept. 6-12, 2026.}
\address{Institute of Theoretical Physics,\\ Faculty of Physics, Astronomy and Applied Computer Science,\\  Jagiellonian University, \\
ul. S. {\L}ojasiewicza 11, 30-348 Krak{\'o}w, Poland.}}

\maketitle

\begin{abstract}
We extend the geometric scaling hypothesis to a $C$-parity odd elastic amplitude and compute the $\rho$ parameters both
for $pp$ and $p\bar{p}$ scattering. These one parameter predictions describe experimental data up to the LHC energies with
high accuracy, but miss the 13~TeV TOTEM result. We show that the introduction of the Odderon allows to describe also 
the TOTEM points and explains the difference of the dip-bump structures of the differential $pp$ and $p\bar{p}$ cross sections.
\end{abstract}


\section{Introduction \label{sec:intro}}

Geometric scaling for $pp$ collisions \cite{DiasDeDeus:1973lde} and the Odderon \cite{Lukaszuk:1973nt} were introduced in 1973, 
more or less at a time when the unexpected rise of hadronic total cross sections with energy was first observed. While the leading 
contribution to elastic hadron scattering at high energies is associated with the Pomeron \cite{Gribov:1961ex,Chew:1961ev}, 
which in QCD is understood as an exchange 
of two reggeized gluons \cite{Low:1975sv}\nocite{Nussinov:1975mw,Kuraev:1977fs}--\cite{Balitsky:1978ic}, 
the Odderon, being odd under charge conjugation, is understood as an exchange of three reggeized gluons 
coupled symmetrically to  a color singlet~\cite{Kwiecinski:1980wb,Bartels:1980pe}. 
Despite its nearly 55-year history, the first experimental indications of the existence of the Odderon emerged only about ten years ago.

We have argued in Ref.~\cite{Praszalowicz:2026yil} that the 2017 TOTEM data at 13 TeV on the total $pp$ cross section \cite{TOTEM:2017asr}
 and on the $\rho^{pp}$ parameter \cite{TOTEM:2017sdy}
(defined as $\rho^{pp}={\rm Re} T_{\rm el}^{pp}(s,t=0)/{\rm Im} T_{\rm el}^{pp}(s,t=0)$) can be reproduced within an approach 
based on geometric scaling (GS), provided that one includes a new $C$-odd amplitude corresponding to the Odderon. 
At lower energies, extending down to the ISR and even below, the data are described very well without the Odderon; 
however, $C$-odd Reggeon exchanges are essential.

Another piece of evidence for the Odderon comes from the TOTEM--D0 analysis of the differential cross section at 1.96~TeV \cite{D0:2020tig}, 
which found a significant difference at the dip between the interpolated $pp$ cross section and the D0 $p\bar{p}$ result. 
The dips in elastic differential cross sections occur at values of $t$ where the imaginary part of the amplitude vanishes, 
so the cross section at the dip is determined entirely by the real part. The $\rho$ parameter and the dip structure are therefore 
two sides of the same coin.

In the present paper, we first recapitulate the results of Ref.~\cite{Praszalowicz:2026yil}, which also provides 
an extensive list of relevant references. We then discuss the differential cross section, showing that 
the difference between the $pp$ and $p\bar{p}$ cross sections at the dip is determined by the $C$-odd amplitude 
and vanishes in the absence of the Odderon contribution.

\section{Geometric scaling and crossing}

Geometric scaling proposed in 1973 \cite{DiasDeDeus:1973lde} for the dimensionless
$C$-even amplitude \cite{DiasdeDeus:1975ybq}
means that 
\begin{equation}
{T}^{+}_{\text{el}}(s,\tau)=isR^{2}(-is)\Phi\Big( \left\vert t \right\vert R^{2}(-is)  \Big) ,
\label{eq:Tplussdef}
\end{equation}
where $R^2(s)$ is so called interaction radius and $\tau=\left\vert t \right\vert  R^{2}(s) $ is a scaling variable. Function $\Phi(\tau)$ is normalized
to 1 at $\tau=0$. An immediate consequence of conjecture (\ref{eq:Tplussdef}) is that the ratio 
of  bump to  dip {\em positions}, $t_{\rm bump}/t_{\rm dip}$, of the differential elastic $pp$
cross sections is energy independent. We confirmed this scaling  for the energies  from the ISR to the LHC \cite{Baldenegro:2024vgg}.
Following \cite{Praszalowicz:2026yil} we assume that also the $C$-odd amplitude exhibits GS
\begin{equation}
{T}_{\text{el}}^{-}(s,\tau)=-sQ^{2}(-is)\Psi \Big(  \left\vert t  \right\vert  Q^{2}(-is) \Big) , 
\label{eq:Tminusdef}%
\end{equation}
where $Q^2(s)$ is a new interaction radius, and $\Psi$  is a new scaling function also normalized to 1 at $\tau=0$. 
Another possibility would be to assume the same scaling
variable in (\ref{eq:Tplussdef}) and (\ref{eq:Tminusdef}), as suggested in \cite{velazquezCorral:2026gex}.
Amplitudes for $pp$ and $p\bar{p}$
take then the following form
\begin{equation}
{T}_{\mathrm{el}}^{pp}={T}_{\mathrm{el}}^{+} 
+{T}_{\mathrm{el}}^{-},
~~~
{T}_{\mathrm{el}}^{p{\bar{p}}}={T}_{\mathrm{el}}^{+}-{T}_{\mathrm{el}}^{-}.
\label{eq:Tppandppbar}
\end{equation}

The advantage of the GS form of  (\ref{eq:Tplussdef}) and   (\ref{eq:Tminusdef}) is that it is now easy to
compute imaginary and real parts of the amplitudes. To this end
one defines rapidity $y$ \cite{DiasdeDeus:1975ybq}
\begin{equation}
-i\left({s}{/{\rm 1\,GeV}^2}\right)=e^{y-i\pi/2}
\end{equation}
and performs expansion in terms of a phase
\begin{equation}
R^{2}(-is) \rightarrow R^{2}\left(  y-i\frac{\pi}{2}\right)  \simeq
R^{2}(y)-i\frac{\pi}{2}\frac{dR^{2}(y)}{dy} \, \label{eq:expansionR}%
\end{equation}
and analogously for $Q^2(-is)$. We obtain \cite{Praszalowicz:2026yil,Praszalowicz:2025twy}
\begin{align}
{T}_{\text{el}}^{+}(s,\tau)  &  =s\left[  iR^{2}\Phi+\frac{\pi}{2}%
\frac{dR^{2}}{dy}\frac{d}{d\tau}\left(  \tau\Phi\right)  \right]  , \nonumber \\
{T}_{\text{el}}^{-}(s,\tau)    & =s\left[  i\frac{\pi}{2}\frac{dQ^{2}%
}{dy}\frac{d}{d\tau}\left(  \tau\Psi\right)  -Q^{2}\Psi\right]\, ,
\label{eq:Tpmfinal}
\end{align}
and%
\begin{align}
T_{\text{el}}^{pp/p\bar{p}}(s,\tau)    = &
s \left[  i\left\{  R^{2}\Phi \pm \frac{\pi}%
{2}\frac{dQ^{2}}{dy}\frac{d}{d\tau}\left(  \tau\Psi\right)  \right\} \right. \nonumber \\
 & \left. +\left\{  \frac{\pi}{2}\frac{dR^{2}}{dy}\frac{d}{d\tau}\left(  \tau
\Phi\right)  \mp Q^{2}\Psi\right\}  \right]  .
\label{eq:Tppandppbarfinal}
\end{align}

From Eqs.\,(\ref{eq:Tppandppbarfinal}) we can compute  total cross sections
\begin{equation}
\sigma_{\mathrm{tot}}^{pp}    =2\left\{  R^{2}+\frac{\pi}{2}\frac
{dQ^{2}}{dy}\right\},  ~~~~
\sigma_{\mathrm{tot}}^{p\bar{p}}    =2\left\{  R^{2}-\frac{\pi}{2}%
\frac{dQ^{2}}{dy}\right\}  , 
\label{eq:sigmatotal}%
\end{equation}
and  $\rho$ parameters
\begin{equation}
\rho^{pp}    =\frac{\frac{\pi}{2}\frac{dR^{2}}{dy}-Q^{2}}{R^{2}+\frac{\pi}%
{2}\frac{dQ^{2}}{dy}},~~~
\rho^{p\bar{p}}   =\frac{\frac{\pi}{2}\frac{dR^{2}}{dy}+Q^{2}}{R^{2}%
-\frac{\pi}{2}\frac{dQ^{2}}{dy}} .
\label{eq:rhos}
\end{equation}

\section{Forward phenomenology and the Odderon}

\begin{figure}[h]
\centering
\includegraphics[width=6.14cm]{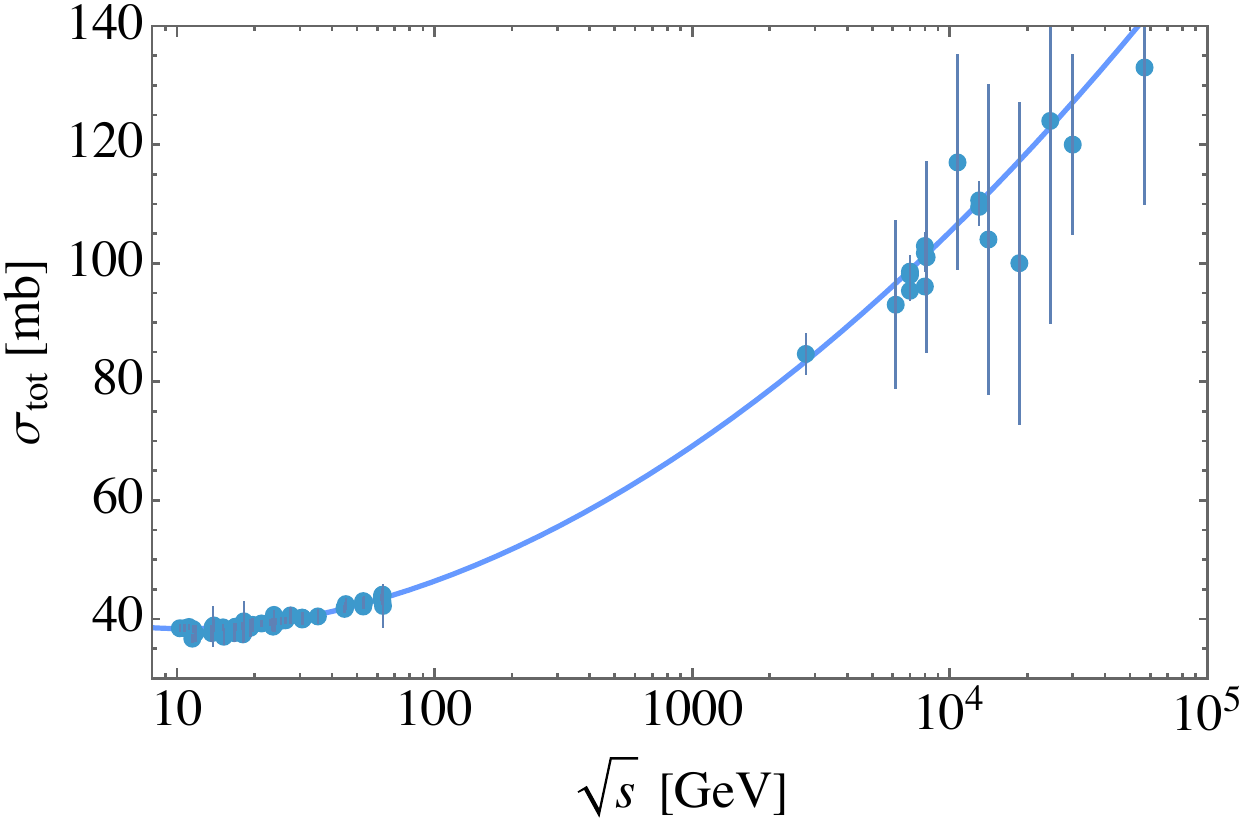}~
\includegraphics[width=6.14cm]{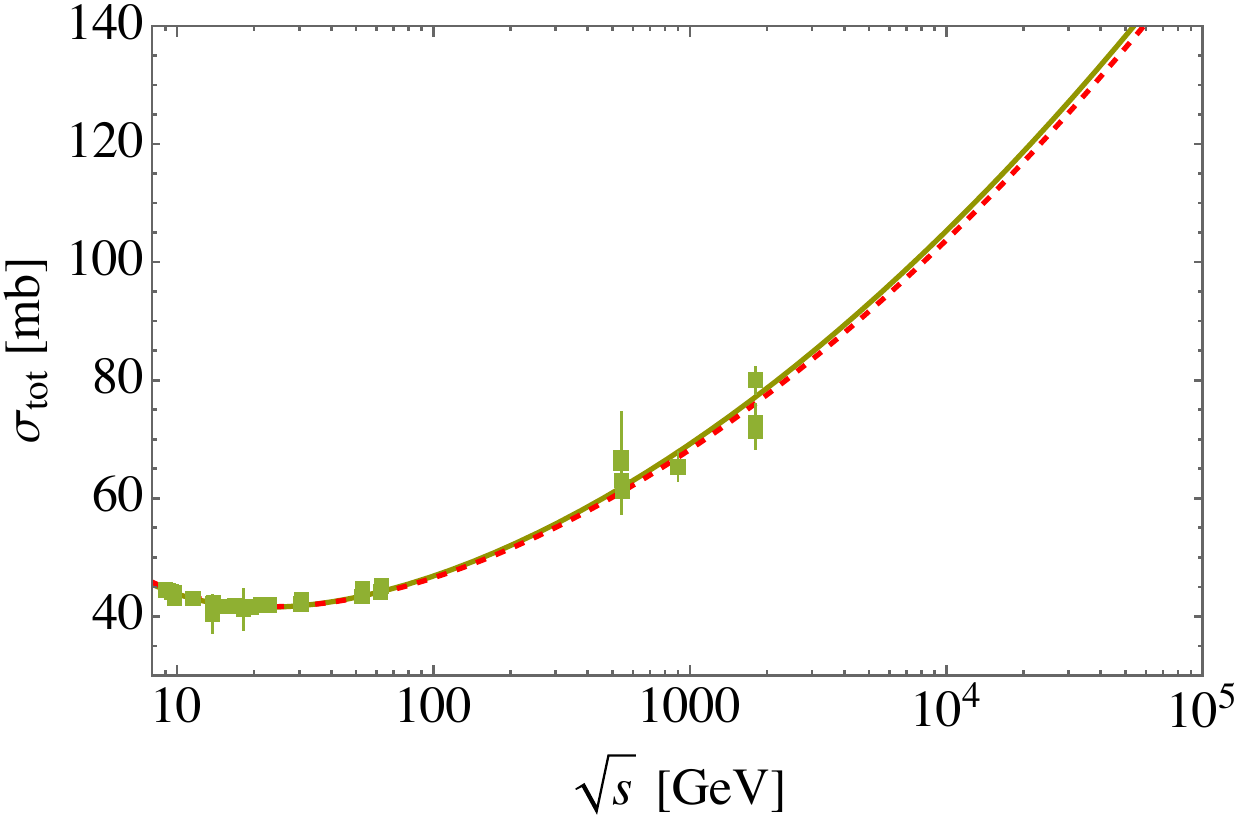}
\vspace{-0.2cm} \caption{Total  cross sections for the COMPETE parametrizations of $pp$ (left)
and $p\bar{p}$ (right) cross-sections --
solid lines. Modified parametrization
(\ref{eq:gauge}) with $f$ given by (\ref{eq:f}) 
 -- dashed red line in the right panel. Data from \cite{data}.}
\label{fig:total}%
\end{figure}
In order constrain interaction radii $R^{2}$ and $Q^{2}$ we use
 the COMPETE parametrization \cite{Cudell:2001pn}
(quoted in \cite{PDG2010}) for the total cross sections shown as solid lines together with data \cite{data} in Fig.~\ref{fig:total}:
\begin{equation}
\sigma_{\mathrm{tot}}^{pp/p\bar{p}}(s)=\underset{=2R^{2}}{\underbrace
{Z+C\ln^{2}\left(  \frac{s}{s_{0}}\right)  +Y_{1}\left(  \frac{s}{s_{1}%
}\right)  ^{-\eta_{1}}}}\mp\underset{=-\pi\,dQ^{2}/dy}{\underbrace
{Y_{2}\left(  \frac{s}{s_{1}}\right)  ^{-\eta_{2}}}}\,,
\label{eq:parPDG}%
\end{equation}
where $Z=35.45$~mb, $C=0.308$~mb, $Y_{1}=42.53$~mb, $Y_{2}=33.34$~mb,
$s_{0}=28.94$~GeV$^{2}$, $s_{1}=1$~GeV$^{2}$ and $\eta_{1}=0.458$, $\eta
_{2}=0.545$. Note that the  two power-like contributions correspond to the
Reggeon exchanges and vanish at high energies,
and that function $Q^2$ is determined up to an integration  constant $q$:
\begin{equation}
Q^{2}(s)    = \frac{1}{\pi\eta_{2}}Y_{2}\left(  \frac{s}{s_{1}}\right)  ^{-\eta_{2}%
}\,+q .
\label{eq:inputs2}
\end{equation}
\begin{figure}[h]
\centering
\includegraphics[width=6.14cm]{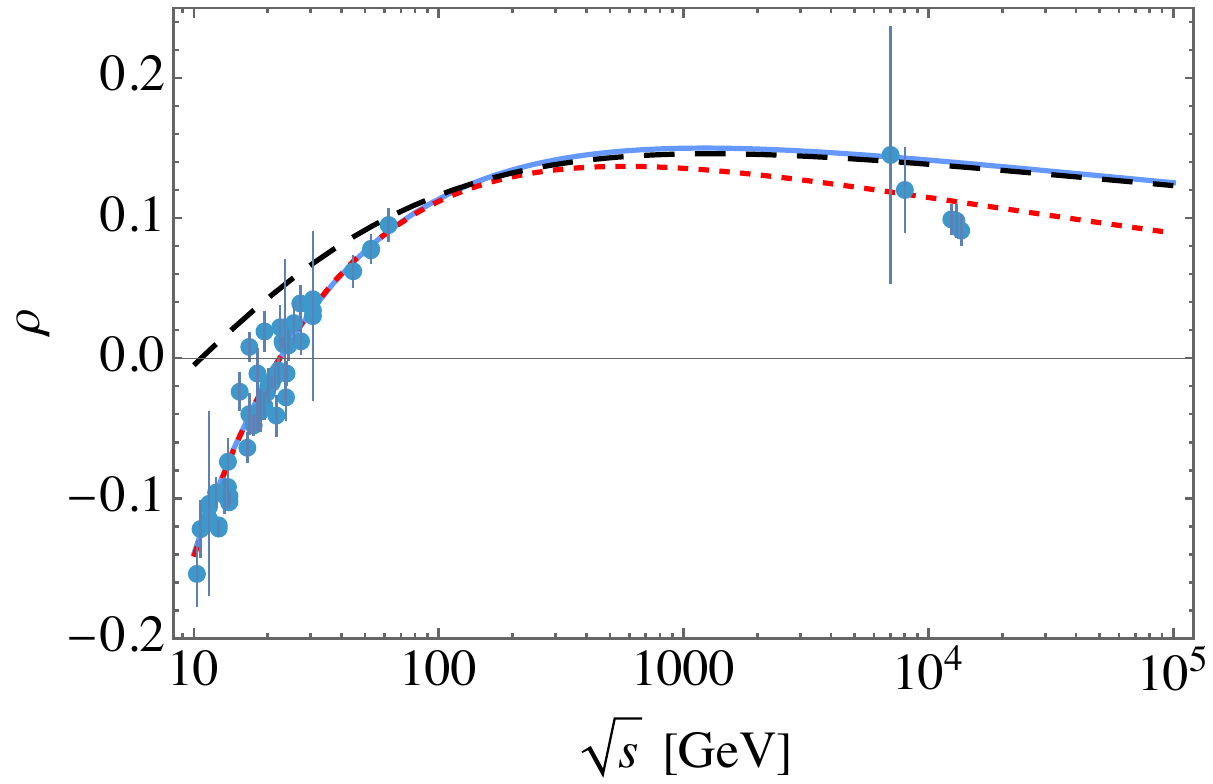}~
\includegraphics[width=6.14cm]{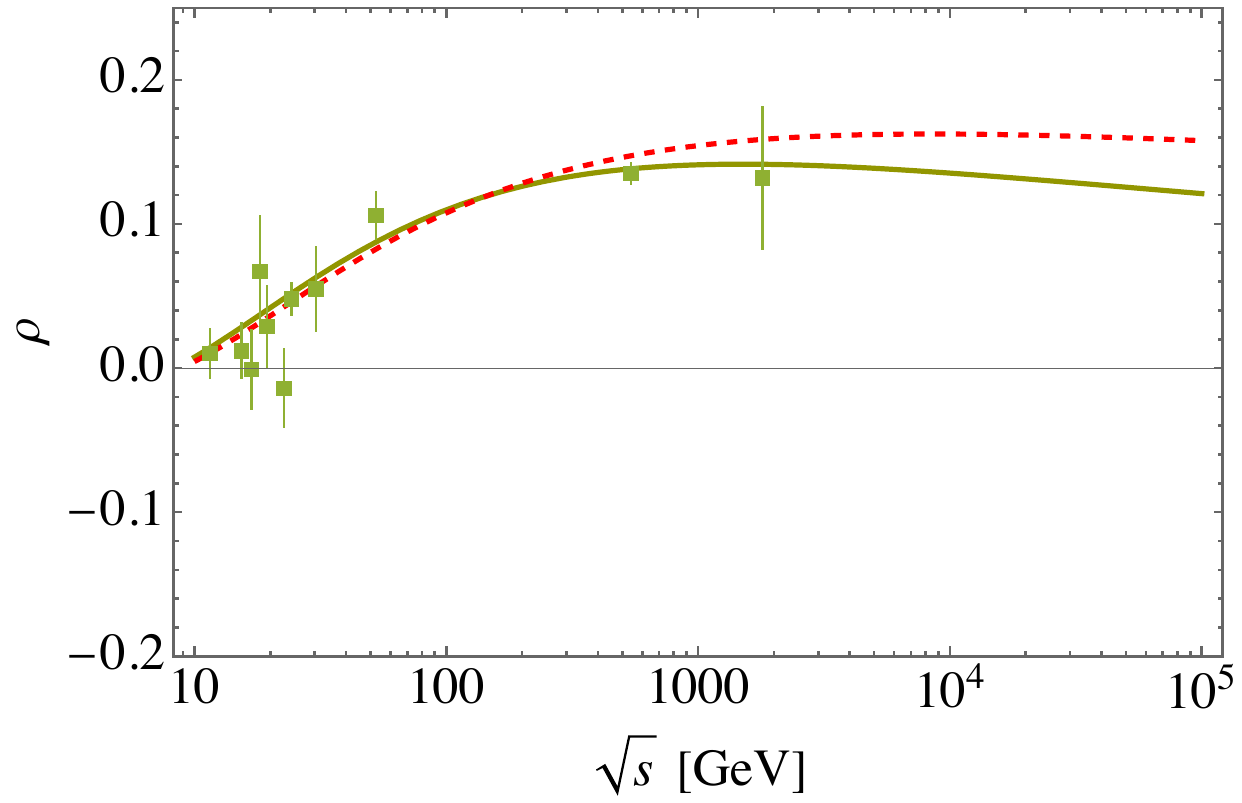}
\vspace{-0.2cm} \caption{\small  $\rho$ parameters for the COMPETE parametrization
corresponding to $q=-0.16$~mb -- solid lines. Modified parametrization
(\ref{eq:gauge}) with $f$ given by (\ref{eq:f})
 -- short dash red lines. 
 Long black dash line  in the left panel -- $\rho^{pp}$ for parametrization without the $C$-odd amplitude \cite{Praszalowicz:2025twy}. 
Left panel: $pp$, right panel: $p\bar{p}$. Data from \cite{data}.}%
\label{fig:modified_rho}%
\end{figure}

Taking parametrization (\ref{eq:parPDG}) as an input, we have absolute predictions for the $\rho$ parameters (\ref{eq:rhos})
up to the integration constant $q$, for which we take  the value $-0.16$~mb, which minimizes the mean squared normalized residual for $\rho$ \cite{Praszalowicz:2026yil}.
The results are plotted as solid lines in Fig.~\ref{fig:modified_rho} together with the prediction from \cite{Praszalowicz:2025twy} (long dash black line), 
where the $C$-odd amplitude was neglected. We  see that taking this amplitude into account is absolutely necessary to correctly describe the 
low energy data. The excellent agreement with experiment below 13~TeV confirms the validity of our conjecture (\ref{eq:Tminusdef}).
However, parametrization (\ref{eq:parPDG}) still misses the 13~TeV TOTEM points.

In order to reproduce the measured value of  $\rho^{pp}$ at 
13 TeV, a modification of $\sigma^{pp}_{\rm tot}$ or/and $\sigma^{p\bar{p}}_{\rm tot}$
parametrizations 
at high energies is necessary. In \cite{Praszalowicz:2026yil} we have introduced the following
modification of the scaling radii
\begin{equation}
\frac{dQ^{2}(y)}{dy}   \rightarrow \frac{dQ^{2}(y)}{dy}-\frac{1}{2\pi}\frac{df(y)}{dy}, ~~
R^{2}(y)   \rightarrow R^{2}(y)+\frac{1}{4}\frac{df(y)}{dy} 
\label{eq:gauge}
\end{equation}
that leaves $\sigma^{pp}_{\rm tot}$ unchanged, but modifies $\sigma^{p\bar{p}}_{\rm tot}$
at high energies, where there is no experimental data. Motivated by the maximal Odderon
conjecture \cite{Lukaszuk:1973nt,Martynov:2017zjz} we have chosen \cite{Praszalowicz:2026yil}
\begin{equation}
f(y)=-A \ln^2 (s/s_3)
\label{eq:f}
\end{equation}
with $A=0.068$~mb, $s_3=1000$~GeV$^2$ and $q=-0.23$~mb. The results for the $\rho$ parameters
are shown in Fig.~\ref{fig:modified_rho} as dashed red lines, and the modification of the $p\bar{p}$ total
cross section is displayed in the right panel of Fig.~\ref{fig:total}.

We see that the COMPETE parametrization correctly reproduces $C$-odd amplitudes
at energies up to the LHC energies below 13~TeV.  However, the Odderon modification is required
in order to describe data up to 13~TeV.
Whether Ansatz (\ref{eq:f}) is the only solution  to solve the TOTEM puzzle is, however, still an open question.

\section{Dip and bump structure of the differential cross sections}

In order to study the dip-bump structures of the differential cross sections, one should, in principle, extract functions
$\Phi$ and $\Psi$ from data. However, one can see if we can, at least qualitatively, explain: (1) whether the bumps and dips
of $pp$ and $p\bar{p}$ cross sections occur at the same values of $t$, (2) whether $pp$ dips are much deeper than in $p\bar{p}$, and
(3) whether bumps in both cases are nearly degenerate.

\begin{figure}[h]
\centering
\includegraphics[width=6cm]{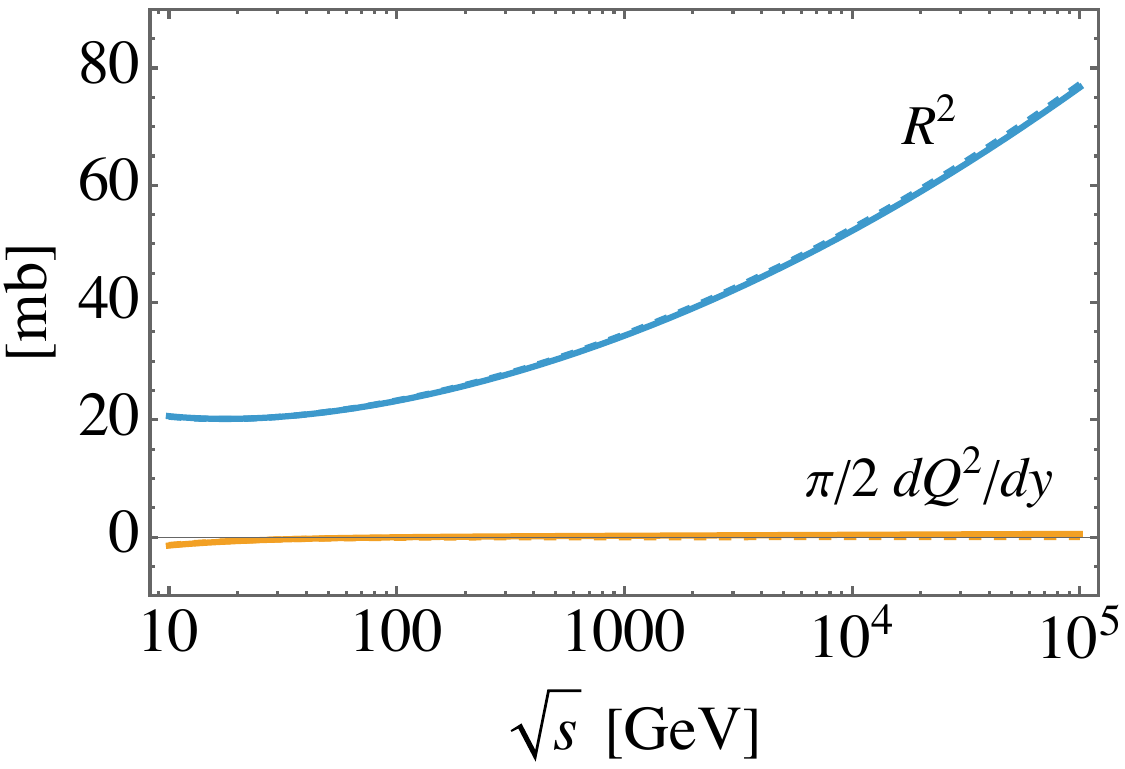}~~
\includegraphics[width=6cm]{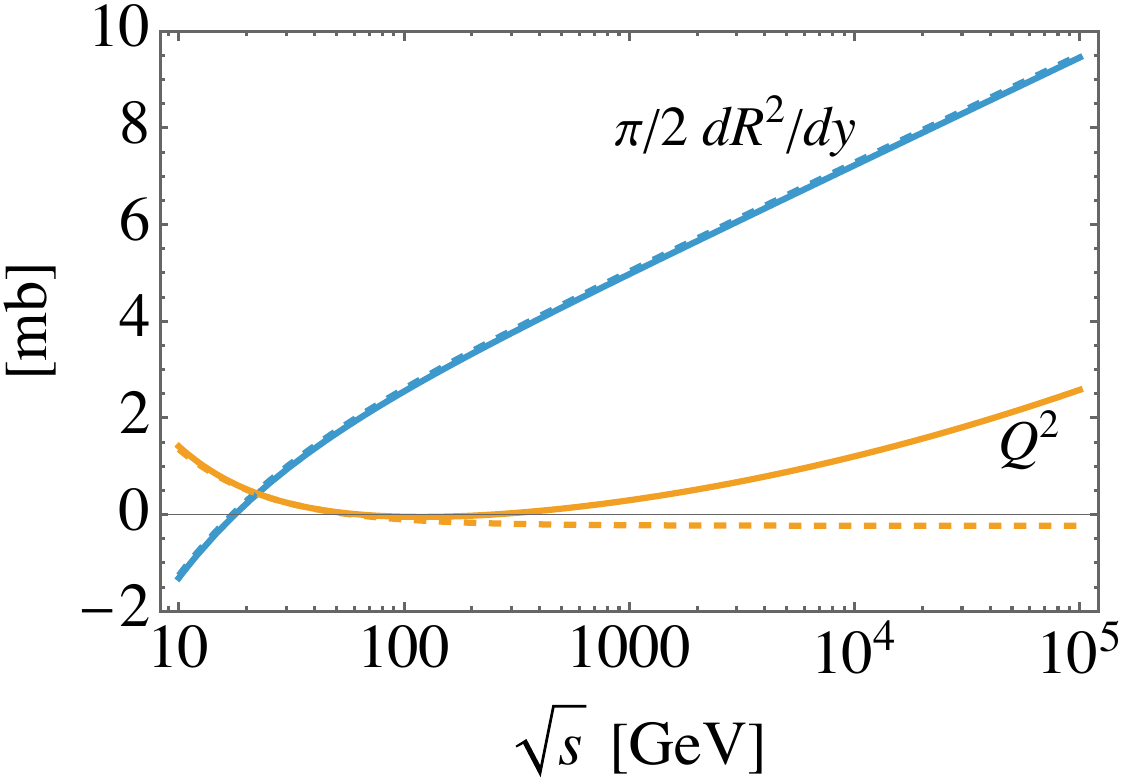}
\vspace{-0.2cm} \caption{\small  
Energy dependent coefficients of the scattering amplitudes (\ref{eq:Tppandppbarfinal}) with $q=-0.23$~mb.
Imaginary part -- left panel, real part -- right panel. Solid lines: with the Odderon contribution ($A=0.068$~mb), dashed line --
without the Odderon ($A=0$).}%
\label{fig:ImRe_of_s}%
\end{figure}

To this end in Fig~\ref{fig:ImRe_of_s} we compare energy dependent coefficients of the scattering amplitudes (\ref{eq:Tppandppbarfinal}).
The results without the Odderon are given by dashed lines. We see that only $Q^2$ is sensitive to the Odderon.
The amplitudes are dominated by the first term in the imaginary part, unless $\Phi$ is very small, and are basically identical
both for $pp$ and $p\bar{p}$ scattering\footnote{Recall that $|\Phi|,\, |\Psi| \leq 1$.}. 
Since the second term of Im$T_{\rm el}^{pp/p\bar{p}}$ is negligibly small, the imaginary part vanishes
 in both cases to a good approximation when $\Phi\simeq 0$, that is for the same value of $\tau$. Note that Im$T_{\rm el}^{pp/p\bar{p}}=0$
 corresponds to a dip.
 
\begin{figure}[h]
\centering
\includegraphics[height=4.25cm]{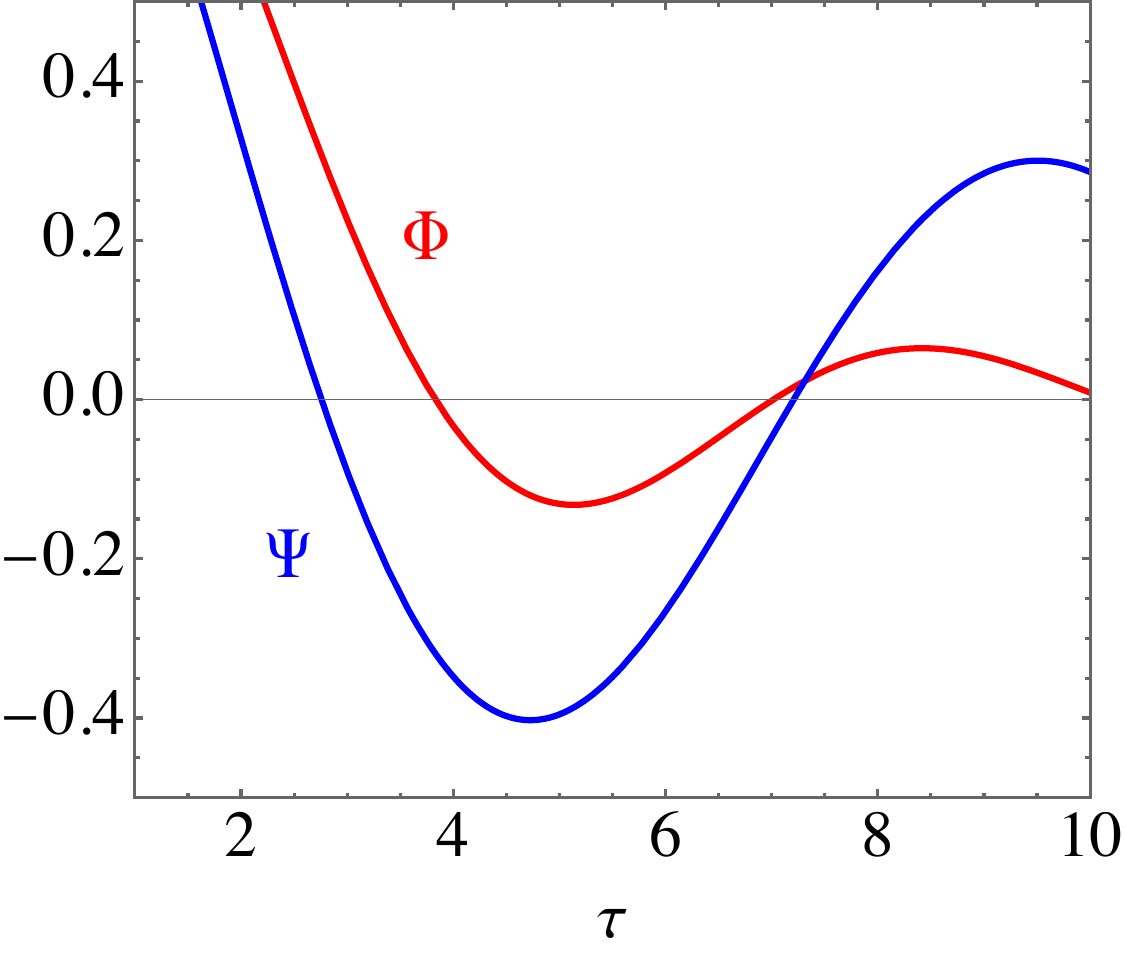}~~
\includegraphics[height=4.3cm]{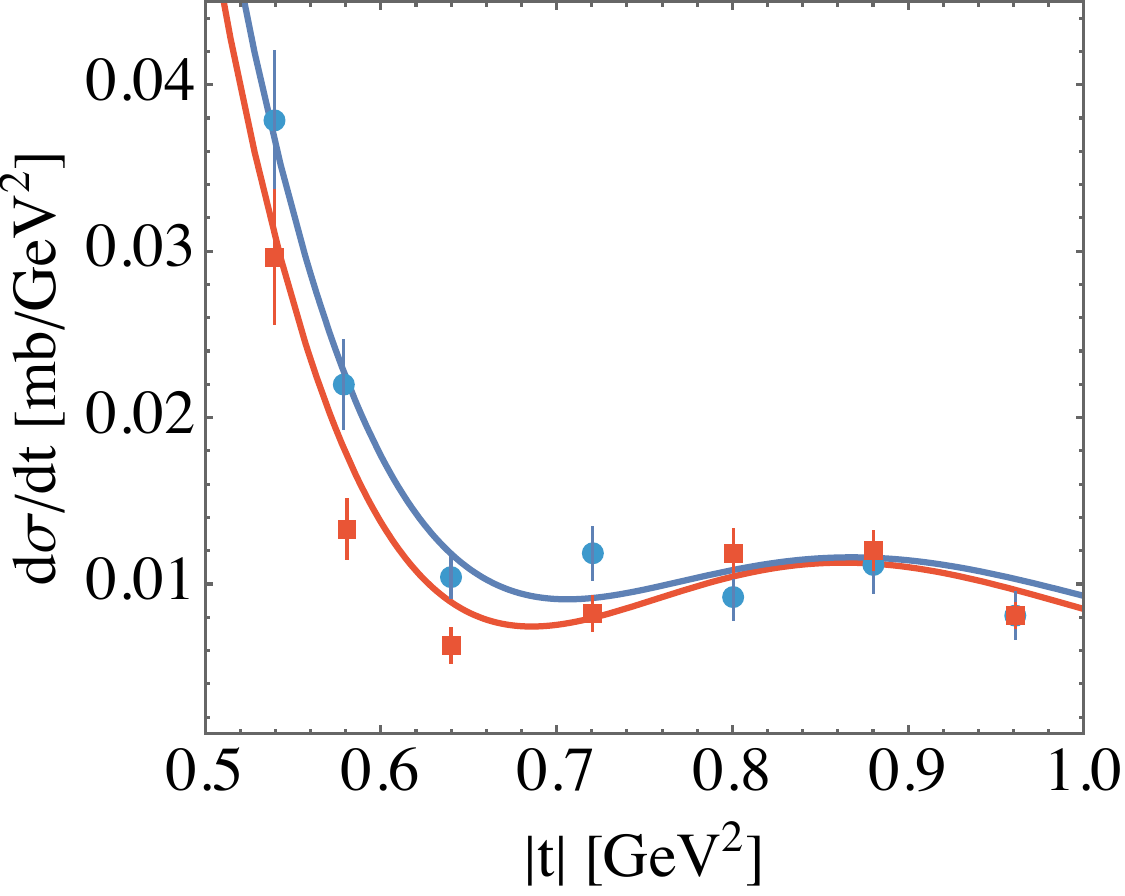}
\vspace{-0.2cm} \caption{\small  Left panel -- scaling functions $\Phi$ and $\Psi$ given by Eq.~(\ref{eq:PhiPsi})
and the resulting cross sections (right panel).
}%
\label{fig:PhiPsi}%
\end{figure}

 Thus the value of the cross sections at the dip is entirely  given by the real part. Here, although  the real part is dominated by the first term
 $\sim dR^2/dy$, the second term $\sim \pm Q^2$ cannot be neglected and contributes to a difference between  $pp$ and $p\bar{p}$ cross sections.
 Numerical value of this splitting, and most importantly the sign, depend on the balance between $d(\tau \Phi)/d \tau$ and $\Psi$. Since at the
 dip $d(\tau \Phi)/d \tau|_{\rm dip} <0$, we see that in order to have $p\bar{p}$ dip shallower than in $pp$, $\Psi|_{\rm dip}$ must be also negative.
 Possible scenario is plotted in the left panel of Fig.~\ref{fig:PhiPsi}. Note that without the Odderon, $Q^2\simeq0$ at high energies and the two
 minima would be degenerate. 
 At the bump $d\Phi/d \tau \simeq 0$ and the tiny differences of the $pp$ and $p\bar{p}$ real parts of the amplitude
 are overshadowed by the dominant imaginary part.
 
 Interestingly, at $\sqrt{s}\simeq 100$~GeV, function $Q^2\simeq 0$ and both minima should be degenerate. 
 There is, unfortunately, no data at this energy range for comparison.
 Below 100~GeV radius $Q^2$ is again nonzero 
 due to the odd $C$ reggeons, and the values of the $pp$ and $p\bar{p}$ cross sections should be again different.
This pattern does not change if we use $|t |R^2(s)$ as a scaling variable of function $\Psi$ (\ref{eq:Tminusdef}).
 
 The question is, whether some "natural" choice of $\Phi$ and $\Psi$ can qualitatively explain the dip and bump structure of the total cross 
 sections at 1.96~TeV provided by the D0 and TOTEM Collaborations \cite{D0:2020tig}? To this end, motivated by the black disk
 amplitude in the $b$ space \cite{Praszalowicz:2025twy}, we have tried the following scaling functions
 \begin{equation}
 \Phi(\tau)=2 \frac{J_1(\tau^{\alpha/2})}{\tau^{\alpha/2}},~~\Psi(\tau)=J_0(\tau^{\beta/2})
 \label{eq:PhiPsi}
 \end{equation}
 with $\alpha=2$ and $\beta=1.73$. These functions, normalized to 1 at $\tau=0$, are plotted in the left panel of Fig.~\ref{fig:PhiPsi} and the
 resulting cross sections are in the right panel, assuming that at 1.96~TeV both scaling variables of $\Phi$ and $\Psi$ 
 are numerically equal.

We see that with these two ad hoc chosen functions one is able to describe qualitatively the experimental data.
Full analysis of the differential cross sections within the present approach is currently under way.

\section{Conclusions and outlook}

In this paper, following \cite{Praszalowicz:2026yil}, we have discussed the  extension of geometric scaling
\cite{DiasDeDeus:1973lde,DiasdeDeus:1975ybq,Baldenegro:2024vgg,Praszalowicz:2025twy} to $C$-parity odd
amplitudes. Extracting  the energy dependent interaction radii $R^2$ and $Q^2$  from the  $pp$ and $p\bar{p}$ total
cross sections we have been able to {\em predict} the $\rho$ parameters with high accuracy up to the highest LHC
energies. In order to reproduce the TOTEM points \cite{TOTEM:2017sdy} we modified $\sigma^{p\bar{p}}_{\rm tot}$ adding the Odderon
contribution (\ref{eq:f}) in such a way that only the total $p\bar{p}$ total cross section is slightly modified above 2~TeV,
where there is no experimental data.

The essential ingredient of the whole approach is  the phase expansion (\ref{eq:expansionR}), which allows to relate
real and imaginary parts of the amplitude up to one free parameter $q$ for the $C$-odd amplitude. This method was already used in 1970-ties \cite{DiasdeDeus:1975ybq}
and was revitalized recently in \cite{Praszalowicz:2025twy}.

Next, we have shown that qualitatively our approach can explain that $pp$ and $p\bar{p}$ cross sections at the same energy $\sqrt{s}$
are different 
at the dips being almost equal elsewhere, and that bumps and eventual dips occur at the same value of $t$.
Further quantitative study of the differential cross sections using the present approach will be presented elsewhere.



\end{document}